\documentstyle[aps,prb,epsf]{revtex}

\begin{document}   

\draft
     

\title{Application of the lattice Green's function for calculating 
the resistance of infinite networks of resistors} 
 
\author{J\'ozsef Cserti}  
\address{E\"otv\"os University, Department of Physics of Complex Systems, 
H-1117 Budapest, P\'azm\'any P\'eter s\'et\'any 1/A , Hungary}   

\maketitle
 
\begin{abstract}  


We calculate the resistance between two arbitrary grid points of 
several infinite lattice structures of resistors by using 
lattice Green's functions.
The resistance for $d$ dimensional hypercubic, rectangular, 
triangular and honeycomb lattices of resistors is discussed in detail.
We give  recurrence formulas for the resistance between arbitrary
lattice points of the square lattice. 
For large separation between nodes we calculate the asymptotic form 
of the resistance for a square lattice and the finite limiting value of 
the resistance for a simple cubic lattice. We point out the relation 
between the resistance of the lattice and the van Hove singularity of 
the tight-binding Hamiltonian. 
Our Green's function method can be applied in a straightforward manner 
to other types of lattice structures and can be useful didactically 
for introducing many concepts used in condensed matter physics.  

\end{abstract}






\section{Introduction} \label{sec:intro}
 
It is an old question to find the resistance between two adjacent grid
points of an infinite square lattice in which all the edges represent 
identical resistances $R$. It is well known that the result is $R/2$, 
and an elegant and elementary solution of the problem is given by 
Aitchison \cite{Aitchison}. 
The electric-circuit theory is discussed in detail in a classic text 
by van der Pol and Bremmer\cite{vanderPol} and they derive 
the resistance between nearby points on the square lattice.
In  Doyle's and Snell's\cite{Doyle} book the connection between 
random walks and electric networks is presented, including many interesting 
results and useful references.   
Recently Venezian \cite{Venezian} 
and Atkinson et.\ al.\ \cite{Atkinson} also studied the 
problem and in these papers the reader can find 
additional references. 
Venezian's method for finding the resistance between two
arbitrary grid points of an infinite square lattice is based on 
the principle of the superposition of current distributions. 
This method was further utilized by Atkinson et.\ al.\ \cite{Atkinson} 
and applied to two dimensional infinite triangular and hexagonal lattices 
as well as infinite cubic and hypercubic lattices. 
In this paper we present an alternative approach using lattice Green's
functions. 
This Green's function method may have several advantages:  
(i) It can be used straightforwardly for more complicated lattice
structures such as body and face centered cubic lattices.
(ii) The results
derived by this method reflect the symmetry of the lattice structures
although they may not be suitable for numerical purposes. However, from these
results one may derive other integral representations of the resistance
between two nodes, which can be used for evaluating the integrals
either algebraically or numerically with high precision. Later in this 
paper we shall give examples for this procedure.    
(iii) From the equation for the Green's function one can, in principle, derive 
some so-called recurrence formulas for the resistances between 
arbitrary grid points of an infinite lattice. 
In this paper such recurrence formulas are derived for an infinite square 
lattice \cite{sajat} 
(for the first time to the best of the author's knowledge).
(iv) In condensed matter physics the application of 
the lattice Green's function has become a very efficient tool. The analytical 
behavior of the lattice Green's function has been extensively studied 
over the past three decades for several lattice
structures. We make use of the knowledge of this analytical behavior
for an infinite cubic lattice and give the asymptotic value of the resistance 
as the separation of the two nodes tends to infinity. 
Throughout this paper we shall utilize the results known in the literature
about the lattice Green's function and also give some important references.  
(v) Finally, our approach for networks of resistors may serve as a 
didactically good example for introducing the Green's function method as
well as many basic concepts such as the Brillouin zone (BZ) used in solid
state physics.
We therefore feel that our Green's function method is of some physical 
interest.

The application of the Green's function proved to be a very effective method
for studying the transport in inhomogeneous conductors and it has been used
successfully by Kirkpatrick \cite{Kirkpatrick} for percolating networks 
of resistors. Our approach for obtaining the resistance of a lattice of 
resistors is closely related to that used by Kirkpatrick.
Economou's book\cite{Economou} gives an excellent introduction to 
the Green's function. A review of the lattice Green's function is given 
by Katsura et.\ al.\ \cite{Katsura}   
The lattice Green's function is also utilized in the theory of 
the Kosterlitz--Thouless--Berezinskii\cite{KT,Kosterlitz,Berez} 
phase transition of the screening of topological defects (vortices).
A review of the latter problem is given by Chaikin et.\ al.\ in
their book \cite{Lubensky}.  
The phase transition in classical two-dimensional lattice Coulomb gases 
has been studied by Lee et.\ al. \cite{Teitel} also by using the 
lattice Green's function in their Monte Carlo simulations. 
The above examples of the applications of the lattice Green's function 
are just a selection of many problems known in the literature. 
A review of the Green's function of the so-called 
tight-binding Hamiltonian (TBH) used for describing the electronic band 
structures of crystal lattices is presented 
in Economou's book \cite{Economou}. The lattice Green's function 
defined in this paper is related to the Green's function of the TBH. 
Below we shall point out that the resistance in a given lattice of 
resistors is related to the Green's function of the TBH at the energy 
at which the density of states is singular. 
This singularity is one of the van Hove singularities of
the density of states \cite{Ziman,Ashcroft,Kittel}.

\section{Hypercubic Lattice}  \label{sec:hyper} 

Consider a $d$ dimensional lattice which consists of all lattice points
specified by position vectors ${\bf r}$ given in the form
\begin{equation}
{\bf r}=l_1{\bf a}_1 +l_2{\bf a}_2 + \cdots + l_d{\bf a}_d,  
\label{d-lattice}
\end{equation}
where ${\bf a}_1,{\bf a}_2,\cdots,{\bf a}_d$ are independent primitive 
translation vectors, and  $l_1, l_2,\cdots,l_d$
range through all integer values (i.e.\ positive and negative integers, as 
well as zero). 
In the case of a $d$ dimensional hypercube all the primitive translation 
vectors have the same magnitude, say $a$, $i.e.,$ 
$|{\bf a}_1|=|{\bf a}_2|=\cdots =|{\bf a}_d| = a$. Here $a$ is the lattice
constant of the $d$ dimensional hypercube.

In the network of resistors we assume that the resistance of 
all the edges of the hypercube is the same, say $R$.    
We wish to find the resistance between the origin and
a given lattice point ${\bf r}_0$ of the infinite hypercube. 
We denote the current that can enter at lattice point 
${\bf r}$ by $I({\bf r})$. However, for measuring the resistance between 
two sites the current has to be zero at all other sites.
Similarly, the potential at site ${\bf r}$ will be denoted by $V({\bf r})$.
Then, at site ${\bf r}$, according to Ohm's and Kirchhoff's laws, 
we may write 
\begin{equation}
I({\bf r})R = \sum_{\bf n} 
\left[V({\bf r})-V({\bf r}+{\bf n})\right],
\label{aram-1}
\end{equation}
where the ${\bf n}$ are the vectors from site ${\bf r}$ 
to its nearest neighbors
(${\bf n} = \pm {\bf a}_i, i=1,\cdots, d$).
The right hand side of Eq.~(\ref{aram-1}) may be expressed by the
so-called {\it lattice Laplacian\/} defined on the hypercubic lattice:
\begin{equation}
\triangle_{({\bf r})}f({\bf r}) = 
\sum_{\bf n} \left[
f({\bf r}+{\bf n})-f({\bf r})
\right].
\label{laplace}
\end{equation}
The above defined lattice Laplacian corresponds to the 
finite-difference representation of the Laplace operator.  
The lattice Laplacian 
$1/a^2\triangle_{({\bf r})}f({\bf r})$ yields the
correct form of the Laplacian in the continuum limit {\it i.e.\/} 
$a\rightarrow 0$.
The lattice Laplacian is widely used to solve partial differential
equations with the finite-difference method\cite{numrec}. 

To measure the resistance between the origin and an arbitrary lattice point 
${\bf r}_0$ we assume that a current $I$ enters at the origin and exits at 
lattice point ${\bf r}_0$.
Therefore, the current is zero at all the lattice points 
except for ${\bf r}=0$ and ${\bf r}_0$, where it is $I$ and $-I$, 
respectively. Thus, Eq.\ (\ref{aram-1}), with the lattice Laplacian, 
can be rewritten as
\begin{equation}
\triangle_{({\bf r})}V({\bf r}) = - I({\bf r}) R,
\label{aram-2}
\end{equation}
where current at lattice point ${\bf r}$ is
\begin{equation}
I({\bf r}) = I \left(\delta_{{\bf r},0} -
\delta_{{\bf r},{\bf r}_0} \right).
\label{aram-3} 
\end{equation}
The resistance between the origin and ${\bf r}_0$ is
\begin{equation} 
R({\bf r}_0)= \frac{V({\bf 0})-V({\bf r}_0)}{I}.
\label{ellen-2}
\end{equation}
To find the resistance we need to solve Eq.\ (\ref{aram-2}). This is a
Poisson-like equation and may be solved by using the lattice 
Green's function:
\begin{equation}
V({\bf r}) = R \sum_{{\bf r}^\prime}
G({\bf r}-{\bf r}^\prime) I({\bf r}^\prime),
\label{V-sol}
\end{equation}
where the lattice Green's function is defined by 
\begin{equation}
\triangle_{({\bf r}^\prime)}G({\bf r}-{\bf r}^\prime) = 
-\delta_{{\bf r},{\bf r}^\prime}.
\label{Green-1}
\end{equation}
Finally, the resistance between
the origin and ${\bf r}_0$ can be expressed by the lattice Green's function. 
Using Eq.\ (\ref{aram-3}) and (\ref{ellen-2}) we obtain
\begin{equation}
R({\bf r}_0) = 2 R [G({\bf 0}) - G({\bf r}_0) ],
\label{R-veg} 
\end{equation}
where we have made use of the fact that the lattice Green's function 
is even, $i.e.$ $G({\bf r})=G(-{\bf r})$.  
Equation (\ref{R-veg}) is our central result for the resistance.

To find the lattice Green's function defined by Eq.\ (\ref{Green-1})
we take periodic boundary conditions at the edges of
the hypercube. Consider a hypercube with $L$ lattice points 
along each side. Thus the total number of sites in the $d$
dimensional hypercube is $L^d$.
Substituting the Fourier transform 
\begin{equation}
G({\bf r}) = \frac{1}{L^d} \sum_{{\bf k} \in {\rm BZ}} 
G({\bf k}) e^{i\bf kr}
\label{Fourier}
\end{equation}
of the lattice Green's function into Eq.\ (\ref{Green-1}), we find 
\begin{equation}
 G({\bf k}) = \frac{1}{\varepsilon({\bf k})},
\label{G-k}
\end{equation} 
for the $d$ dimensional hypercube
where we have defined
\begin{equation}
\varepsilon({\bf k}) =  
2\sum_{i=1}^{d} \, \left(1-\cos{\bf k}{\bf a}_i \right).
\end{equation}
Owing to the periodic boundary conditions, the wave vector 
${\bf k}$ in Eq.\ (\ref{G-k}) is limited to the first Brillouin
zone and is given by 
\begin{equation}
{\bf k} = \frac{m_1}{L}{\bf b}_1 + \frac{m_2}{L}{\bf b}_2 
+ \cdots + \frac{m_d}{L} {\bf b}_d,
\label{k-vector}
\end{equation}
where $m_1,m_2,\cdots\,m_d$ are integers such that $-L/2 \leq m_i \leq  L/2$ 
for $i=1,2,\cdots,d$, and ${\bf b}_j$ are the reciprocal lattice vectors
defined by ${\bf a}_i {\bf b}_j = 2\pi \delta_{ij}$, $i,j=1,2,\cdots,d$. 
Here we assumed that $L$ is an even integer, which will be irrelevant in
the limit $L \rightarrow \infty$.
The mathematical description of the crystal lattice and the concept of 
the Brillouin zone can be found in many books on 
solid state physics\cite{Lubensky,Ashcroft,Ziman,Kittel}.
 
Finally, the lattice Green's function takes the form  
\begin{equation}
G({\bf r}) = \frac{1}{L^d} 
\sum_{{\bf k} \in {\rm BZ}} \, 
\frac{e^{i{\bf k}{\bf r}}}{\varepsilon({\bf k})}.
\end{equation}
If we take the limit $L \rightarrow \infty$ then the discrete summation
over ${\bf k}$ can be substituted by an integral 
\cite{Ashcroft,Lubensky}:
\begin{equation}
\frac{1}{L^d} \sum_{{\bf k} \in {\rm BZ}}  \rightarrow 
v_0 \int_{{\bf k} \in {\rm BZ}} \, \frac{d^d {\bf k}}{{(2\pi)}^d},
\label{sum-int}
\end{equation}  
where $v_0=a^d$ is the volume of the unit cell of the $d$ 
dimensional hypercube.
Thus the lattice Green's function is
\begin{equation}
G({\bf r}) = v_0\int_{{\bf k} \in {\rm BZ}} \, 
\frac{d^d {\bf k}}{{(2\pi)}^d}\, 
\frac{e^{i{\bf k}{\bf r}}}{\varepsilon({\bf k})}.
\label{G-veg}
\end{equation}

Using Eqs.\ (\ref{R-veg}) and (\ref{G-veg}) in $d$ dimensions the 
resistance between the origin and lattice point ${\bf r}_0$ is
\begin{equation}
R({\bf r}_0)=
2 R v_0 \int_{{\bf k} \in {\rm BZ}} \, \frac{d^d {\bf k}}{{(2\pi)}^d}\,
\frac{1-e^{i{\bf k}{\bf r}_0}}{\varepsilon({\bf k})}.
\label{R-hyper-1}
\end{equation}
The above result can be simplified if we specify the lattice point as
${\bf r}_0=l_1{\bf a}_1 +l_2{\bf a}_2 + \cdots + l_d{\bf a}_d$: 
\begin{equation}
R(l_1,l_2, \cdots, l_d)= R 
\int_{-\pi}^{\pi} \frac{d x_1}{2\pi}
\cdots 
\int_{-\pi}^{\pi} \frac{d x_d}{2\pi}\, \, \,  
\frac{1-e^{i(l_1 x_1+ \cdots +l_d x_d)}}
{\sum_{i=1}^d ( 1-\cos x_i)}.
\label{R-hyper-2}
\end{equation}
From this final expression of the 
resistance one can see that the resistance does not depend on the 
angles between the unit vectors 
${\bf a}_1,{\bf a}_2,\cdots,{\bf a}_d$. 
Physically this means that the hypercube can be deformed 
without the change of the resistance between any two lattice points. 
The resistance in topologically equivalent lattices is the same.   
For further references we also give the lattice Green's function for
a $d$ dimensional hypercube:
\begin{equation}
G(l_1,l_2, \cdots, l_d)=  
\int_{-\pi}^{\pi} \frac{d x_1}{2\pi}
\cdots 
\int_{-\pi}^{\pi} \frac{d x_d}{2\pi}\, \, \,  
\frac{e^{i(l_1 x_1+ \cdots +l_d x_d)}}
{2\sum_{i=1}^d ( 1-\cos x_i)}.
\label{G-hypercube}
\end{equation}

\subsection{Conducting medium, continuum limit}

The resistance in an infinite conducting medium can be obtained by
taking the limit as the lattice constant $a$ tends to zero 
in  Eq.\ (\ref{R-hyper-1}). 
Denoting the electrical conductivity of the medium 
by $\sigma$, the resistance of a $d$ dimensional hypercube with lattice
constant $a$, according to Ohm's law, is given by
\begin{equation}
R=\frac{1}{\sigma a^{d-2}}.
\label{R-medium}
\end{equation}

Using the approximation 
$\varepsilon({\bf k})\approx {\bf k}^2 a^2 $ 
for $|{\bf a}_i|=a \rightarrow 0$, Eq.\ (\ref{R-hyper-1}) can easily be 
reduced to 
\begin{equation}
R({\bf r}_0)= \frac{2}{\sigma} 
\int_{{\bf k} \in {\rm BZ}} \, \frac{d^d {\bf k}}{{(2\pi)}^d}\,
\frac{1-e^{i{\bf k}{\bf r}_0}}{{\bf k}^2}.
\end{equation}
The same result for the resistance of a conducting medium is
given in Chaikin's book\cite{Lubensky}.

\subsection{Linear chain, $d=1$}

Consider a linear chain of identical resistors $R$. The resistance between 
the origin and site $l$ can be obtained by taking $d=1$ in the general 
result given in  Eq.\ (\ref{R-hyper-2}):
\begin{equation}
R(l)= R 
\int_{-\pi}^{\pi} \frac{d x}{2\pi} \, \, \,    
\frac{1-e^{il x}}{1-\cos x}.
\label{R-d1}
\end{equation}
The integral can be evaluated by the method of residues\cite{Arfken}
and gives the following very simple result:
\begin{equation}
R(l)= R l.
\end{equation}
This can be interpreted as the resistance of $l$
resistances $R$ in series. The current flows only between the two
sites separated by a finite distance. The two semi-infinite segments of the
chain do not affect the resistance.

\subsection{Square lattice, $d=2$}

Using Eq.\ (\ref{R-hyper-2}), the resistance in two dimensions
between the origin and ${\bf r}_0=l_1{\bf a}_1 +l_2{\bf a}_2$ is
\begin{eqnarray}
R(l_1,l_2)  =  R \int_{-\pi}^{\pi} \frac{d x_1}{2\pi}
\int_{-\pi}^{\pi} \frac{d x_2}{2\pi}\, \, \,  
\frac{1-e^{il_1 x_1+il_2 x_2}}{2-\cos x_1 - \cos x_2}
\label{R-d2-a} \\[1ex]
 =  R \int_{-\pi}^{\pi} \frac{d x_1}{2\pi}
\int_{-\pi}^{\pi} \frac{d x_2}{2\pi}\, \, \,  
\frac{1-\cos\left(l_1 x_1+l_2 x_2\right)}{2-\cos x_1 - \cos x_2}.
\label{R-d2-b}
\end{eqnarray}
The resistance between two adjacent lattice sites can easily be obtained
from the above expression without evaluating the integrals in 
Eq.\ (\ref{R-d2-a}) or (\ref{R-d2-b}). 
Note that interchanging $l_1$ and $l_2$ in Eq.\ (\ref{R-d2-b})  
does not change the resistance, $i.e.,$ $R(l_1,l_2)=R(l_2,l_1)$. 
This is consistent with the symmetry of the lattice. 
Then, from Eq.\ (\ref{R-d2-b}), the sum of $R(0,1)$ and $R(1,0)$ yields 
\begin{equation}
R(0,1)+R(1,0)= R 
\int_{-\pi}^{\pi} \frac{d x_1}{2\pi}
\int_{-\pi}^{\pi} \frac{d x_2}{2\pi}=R.
\label{R-d2-c}
\end{equation}
Thus $R(0,1)=R(1,0)=R/2$. The resistance between two adjacent lattice
sites is  $R/2$, which is a well known result \cite{Aitchison,Venezian}.

In general the integrals in Eq.\ (\ref{R-d2-a}) have to be evaluated
numerically. It is shown in Appendix \ref{appendix-1} how one integral can
be performed in Eq.\ (\ref{R-d2-a}). We found the same result as 
Venezian\cite{Venezian} and Atkinson\cite{Atkinson}:
\begin{equation}
R(l_1,l_2)= R \int_{0}^{\pi} \frac{d y}{\pi} \,\,
\frac{1-e^{-\left|l_1\right|s}\cos l_2y}{\sinh s},
\label{R-d2-one_int}
\end{equation}
where
\begin{equation}
\cosh s = 2-\cos y.
\end{equation}
It turns out that this expression is more stable numerically 
than Eq.\ (\ref{R-d2-b}). As an application of the above formula
one can calculate the resistance between second nearest neighbors exactly. 
After some algebra one finds 
\begin{equation}
R(1,1)=R \int_{0}^{\pi} \frac{d y}{\pi} \,\,
\frac{{(1-\cos y)}^2}{\sqrt{{(2-\cos y)}^2-1}}=\frac{2}{\pi}R.
\end{equation}

The energy dependent lattice Green's function of the tight-binding
Hamiltonian for a square lattice is given by
\begin{equation}
G(E;l_1,l_2)= \int_{-\pi}^{\pi} \frac{d x_1}{2\pi}
\int_{-\pi}^{\pi} \frac{d x_2}{2\pi}
\frac{\cos\left(l_1 x_1+l_2 x_2 \right)}
{E-\cos x_1 - \cos x_2}
\label{G_E-d2}
\end{equation}
[see Eq.\ (5.31) in Economou's book\cite{Economou}].
This is a generalization of our Green's function  by introducing a 
new variable $E$ instead of the value 2 in the denominator in 
Eq.\ (\ref{G-hypercube}) for $d=2$. Note that a factor 2 
appearing in the denominator of our Green's function 
in Eq.\ (\ref{G-hypercube}) 
is missing in Eq.\ (\ref{G_E-d2}). This is related to the fact that in 
the Schr\"odinger equation the Laplacian is multiplied by a factor of
$1/2$ while in our case the Laplace equation is solved. 
Comparing Eqs.\ (\ref{G-hypercube}) (for $d=2$) and (\ref{G_E-d2}) 
one can see that the resistance is
\begin{equation} 
R(l_1,l_2)=R\left[G(2;0,0)-G(2;l_1,l_2)\right].
\label{vanHove-sq}
\end{equation}
Based on the equation for the Green's function Morita \cite{Morita-sq} 
derived the recurrence formulas for the Green's function $G(E;l_1,l_2)$ 
for an infinite square lattice (see Eqs.\ (3.8) and
(4.2)-(4.4) in Morita's paper\cite{Morita-sq}). Applying Morita's results 
(with $E=2$)
to the resistance given in Eq.\ (\ref{vanHove-sq})
we obtained the following recurrence formulas for the resistance: 
\begin{eqnarray}
R(m+1,m+1) &=& \frac{4m}{2m+1}\, R(m,m)-\frac{2m-1}{2m+1}\, R(m-1,m-1),
\nonumber \\
R(m+1,m) &=& 2 R(m,m)-R(m,m-1), 
\nonumber \\
R(m+1,0) &=& 4 R(m,0)-R(m-1,0)-2R(m,1),
\nonumber \\
R(m+1,p) &=& 4 R(m,p)-R(m-1,p)-R(m,p+1)- R(m,p-1) 
\,\,\,\, {\rm if} \,\,\,\, 0< p < m.
\label{recur-sq}
\end{eqnarray} 
We have seen that $R(1,0)=R/2$ and $R(1,1)=2R/\pi.$  
Since we know the exact values of $R(1,0)$ and $R(1,1)$ (obviously
$R(0,0)=0$) one can calculate 
the resistance exactly for arbitrary nodes by using the above 
given recurrence formulas.
This way we obtained the same results as 
Atkinson et.\ al.\ \cite{Atkinson} using Mathematica.
The advantages of our recurrence relations are that they provide a 
new, very simple and effective tool to calculate the resistance 
between arbitrary nodes on a square lattice.
We note that van der Pol and Bremmer\cite{vanderPol} also gave the exact 
values of the resistance for nearby points in a square lattice using a
different approach. 

It is interesting to see the asymptotic form of the resistance for 
large values of $l_1$ or/and $l_2$. In Appendix \ref{appendix-2} 
we derive the asymptotic form of the lattice Green's function 
for a square lattice [see Eq.\ (\ref{G-asym-veg})]. 
Inserting  Eq.\ (\ref{G-asym-veg}) into the general result of the 
resistance given in Eq.\ (\ref{R-veg}) the asymptotic form of the 
resistance is
\begin{equation}
R(l_1,l_2)= \frac{R}{\pi} \left( \ln \sqrt{l_1^2+l_2^2} +\gamma + \frac{\ln
8}{2} \right),
\label{R-d2-asym}
\end{equation}
where $\gamma = 0.5772\dots$ is the Euler-Mascheroni
constant\cite{Arfken}. 
The same result was obtained by Venezian\cite{Venezian} 
except that we got an exact value of the numerical constant 
in Eq.\ (\ref{R-d2-asym}) whereas it was numerically approximated 
in Venezian's paper. The resistance is logarithmically
divergent for large values of $l_1$ and $l_2$. 
A similar behavior was found for conducting medium 
by Chaikin et.\ al.\ in their book\cite{Lubensky}.
In the theory of 
the Kosterlitz--Thouless--Berezinskii\cite{KT,Kosterlitz,Berez} 
phase transition of the screening of topological defects (vortices) 
the same asymptotic form of the Green's function (as given in 
Eq.\ (\ref{G-asym-veg})) has been used for a square lattice \cite{Kosterlitz}. 
Finally, we note that Doyle and Snell showed in their book\cite{Doyle} 
(pp.\ 122--123) that the 
resistance goes to infinity as the separation between nodes tends to
infinity but the asymptotic form was not derived. 
  
\subsection{Simple cubic lattice, $d=3$}

In three dimensions  the resistance between the origin and a lattice point
${\bf r}_0=l_1{\bf a}_1 +l_2{\bf a}_2 +l_3{\bf a}_3$ can be obtained 
from Eq.\ (\ref{R-hyper-2}):
\begin{eqnarray}
R(l_1,l_2,l_3) & = & R \int_{-\pi}^{\pi} \frac{d x_1}{2\pi}
\int_{-\pi}^{\pi} \frac{d x_2}{2\pi}
\int_{-\pi}^{\pi} \frac{d x_3}{2\pi}\, \, \,  
\frac{1-e^{il_1 x_1+il_2 x_2+il_3 x_3}}{3-\cos x_1 - \cos x_2-\cos x_3}
\label{R-d3-a} \\[1ex]
& = & R\int_{-\pi}^{\pi} \frac{d x_1}{2\pi}
\int_{-\pi}^{\pi} \frac{d x_2}{2\pi}
\int_{-\pi}^{\pi} \frac{d x_3}{2\pi}\, \, \,  
\frac{1-\cos\left(l_1 x_1+l_2 x_2 + l_3 x_3\right)}
{3-\cos x_1 - \cos x_2 -\cos x_3}.
\label{R-d3-b}
\end{eqnarray}
Similarly to the case of a square lattice, the exact value of 
the resistance between two adjacent lattice sites can be obtained 
from the above expression. 
Clearly, from Eq.\ (\ref{R-d3-b}) (and for symmetry reasons, too) 
$R(1,0,0)=R(0,1,0)=R(0,0,1)$ and 
\begin{equation}
R(1,0,0)+R(0,1,0)+R(0,0,1)=
R\int_{-\pi}^{\pi} \frac{d x_1}{2\pi}
\int_{-\pi}^{\pi} \frac{d x_2}{2\pi}
\int_{-\pi}^{\pi} \frac{d x_3}{2\pi}=R. 
\end{equation}
Therefore, the resistance between adjacent sites is $R/3$. 
It is easy to show much in the same way that in a $d$ dimensional 
hypercube the resistance between adjacent sites is $R/d$.

Unlike in a square lattice, in a simple cubic lattice the resistance 
does not diverge as the separation of the entering and exiting sites
increases, but tends to a finite value.
One can write for the resistance 
$R(l_1,l_2,l_3)=2\left[G(0,0,0)-G(l_1,l_2,l_3)\right]$ where 
$G(l_1,l_2,l_3)$ is given in Eq.\ (\ref{G-hypercube}) for $d=3.$ 
It is well known from the theory of Fourier series (Riemann's lemma) 
that $\lim_{p\rightarrow \infty} 
\int_a^b \, dx \, \varphi(x) \cos px \rightarrow 0$ for any 
integrable function $\varphi(x).$
Hence, $G(l_1,l_2,l_3)\rightarrow 0$ (which indeed corresponds to the
boundary condition of the Green's function at infinity) and thus,  
$R(l_1,l_2,l_3) \rightarrow 2G(0,0,0)$ 
when any of the $l_1,l_2,l_3 \rightarrow \infty.$
The value of $G(0,0,0)$ was evaluated for the first time 
by Watson \cite{Watson} and subsequently by Joyce \cite{Joyce} in a closed
form in terms of elliptic integrals. 
The following exact result was
found
\begin{equation}
2G(0,0,0)={\left(\frac{2}{\pi}\right)}^2
\left(18+12\sqrt{2}-10\sqrt{3}-7\sqrt{6}\right)
{\left[{\bf K}(k_0)\right]}^2 = 0.505462\dots,
\label{exact-kocka}
\end{equation} 
where $k_0=(2-\sqrt{3})(\sqrt{3}-\sqrt{2})$ and 
\begin{equation}
{\bf K}(k)=\int_0^{\pi/2}\, d\theta \, \frac{1}{\sqrt{1-k^2\sin^2 \theta}}
\label{elliptic}
\end{equation}
is the complete elliptic integral of the first kind. 
It is worth mentioning that a simpler result was obtained by 
Glasser et.\ al.\ \cite{Glasser1} 
(see also Doyle's and Snell's book\cite{Doyle}) 
who calculated the integrals 
in terms of gamma functions: 
\begin{equation}
2G(0,0,0)=\frac{\sqrt{3}-1}{96\pi^3}\Gamma^2(1/24)\Gamma^2(11/24).
\label{exact-kocka-Glasser}
\end{equation} 
Thus, the resistance 
in units of $R$ for a 
simple cubic lattice tends to the finite value 0.505462$\dots$ when the
separation between the entering and exiting sites tends to infinity.

\begin{figure}
{\centerline{\leavevmode \epsfxsize=8.5cm \epsffile{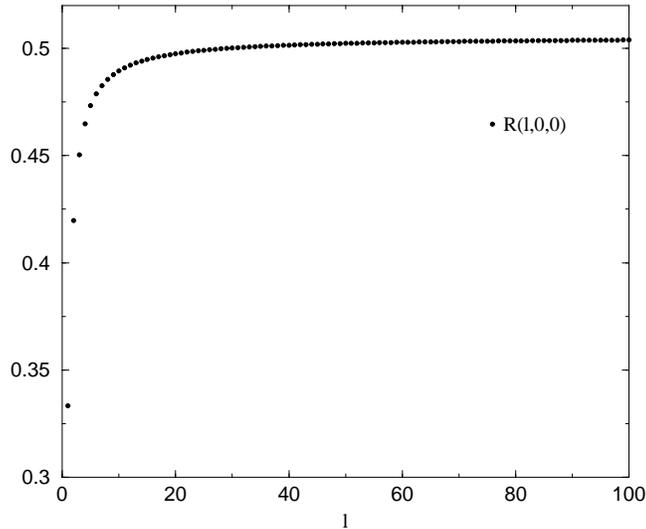 }}}
\caption{The resistance $R(l,0,0)$ in units of $R$ along an axis for 
a simple cubic lattice.
\label{R-gorbe-kocka}}
\end{figure}

For finite separations the formula for the resistance given in 
Eq.\ (\ref{R-d3-b}) is not suitable for numerical purposes since the
integrals converge slowly with increasing the 
number of mesh points. 
Similarly to the case of a square lattice, we can use the energy 
dependent lattice Green's function of the tight-binding
Hamiltonian defined for a simple cubic lattice as 
\begin{equation}
G(E;l_1,l_2,l_3)= \int_{-\pi}^{\pi} \frac{d x_1}{2\pi}
\int_{-\pi}^{\pi} \frac{d x_2}{2\pi}
\int_{-\pi}^{\pi} \frac{d x_3}{2\pi}\, \, \,  
\frac{\cos\left(l_1 x_1+l_2 x_2 + l_3 x_3\right)}
{E-\cos x_1 - \cos x_2 -\cos x_3}
\label{G_E-d3}
\end{equation}
[see Eq.\ (5.49) in Economou's book \cite{Economou}].
This is a generalization of our Green's function by introducing a 
new variable $E$ instead of the value 3 in the denominator in 
Eq.\ (\ref{G-hypercube}) for $d=3$. Note that the reason for the missing 
factor of 2 in Eq.\ (\ref{G_E-d3}) is the same as that explained in the
previous section.   
Comparison of Eqs.\ (\ref{G-hypercube}) (for $d=3$) and (\ref{G_E-d3}) 
yields
\begin{equation} 
R(l_1,l_2,l_3)=R\left[G(3;0,0,0)-G(3;l_1,l_2,l_3)\right].
\label{vanHove}
\end{equation}
The analytic behavior of the lattice Green's function 
$G(E;l_1,l_2,l_3)$ has been extensively studied over the past three decades.
Numerical values of $G(E;l_1,l_2,l_3)$ were given by Koster et.\ al., 
\cite{Koster} and Wolfram et.\ al., \cite{Wolfram} 
by using the following representation of the Green's function 
for a simple cubic lattice:
\begin{equation}
G(E\geq 3;l,m,n)= \int_0^\infty \, dt \, 
e^{-Et}I_{l}(t)I_{m}(t)I_{n}(t),
\end{equation}
where $I_{l}$ is the modified Bessel function of order $l$. However, it
turns out that  another representation of the
lattice Green's functions is more suitable for numerical calculations. 
For $3\leq E$ the following form of the Green's function along an axis 
was given in terms of complete elliptic integrals of the first kind 
by Horiguchi \cite{Horiguchi}
\begin{equation}
G(E;l,0,0)= \frac{1}{\pi^2}\int_0^\pi \, dx \, {\bf K}(k) k \cos lx,
\label{Hori}
\end{equation}
where 
\begin{equation}
k=\frac{2}{E-\cos x}
\label{Hori-k}
\end{equation}
and ${\bf K}(k)$ is defined in Eq.\ (\ref{elliptic}).
In Fig.\ \ref{R-gorbe-kocka} we plotted the numerical values of 
the resistance using Eq.\ (\ref{Hori}) for $E=3$ and $1\leq l \leq 100$.
It is seen from the figure that the resistance tends rapidly to 
its asymptotic value given in Eq.\ (\ref{exact-kocka}). 
We would mention that Glasser et.\ al.\ \cite{Glasser2} gave other 
useful integral representations of the lattice Green's function 
for the hypercubic lattice for arbitrary dimension $d$.

It was shown by Joyce \cite{Joyce} that the function $G(E;0,0,0)$ can be 
expressed in the form of a product of two complete elliptic 
integrals of the first kind. Horiguchi\cite{Horiguchi} and later on 
Morita\cite{Morita} obtained recurrence 
relations for the function $G(E;l_1,l_2,l_3)$ of the simple cubic 
lattice for an arbitrary site $l_1,l_2,l_3$  in terms only of 
$G(E;0,0,0)$, $G(E;2,0,0)$ and $G(E;3,0,0)$. The last two Green's functions 
were expressed by Horiguchi and Morita \cite{Hori-Morita} in closed form in
terms of complete elliptic integrals. From these results one can calculate
the Green's function for an arbitrary lattice point.
As an example, one of the recurrence formulas we obtained 
from the recurrence formula of the Green's function (at $E=3$) 
given by Horiguchi\cite{Horiguchi} is 
\begin{equation}
R(m,1,0)=-\frac{1}{4}\, R(m-1,0,0)+\frac{3}{2}\, R(m,0,0)
-\frac{1}{4}\, R(m+1,0,0).
\end{equation}  
Together with other recurrence formulas given 
in Horiguchi's and Morita's paper 
one can find $R(l_1,l_2,l_3)$ for arbitrary values of $l_1,l_2,l_3$.
The numerical values of the resistance for small 
$l_1,l_2,l_3$ are listed in Table I. 

\newpage
\begin{table}[hbt]
\begin{tabular}{cc}  \\
$l_1,l_2,l_3$ & $R(l_1,l_2,l_3)/R$ \\[1ex]  \hline 
1,0,0 &  $1/3$ \\   
1,1,0 &  0.395079 \\   
1,1,1 &  0.418305 \\   
2,0,0 &  0.419683 \\   
2,1,0 &  0.433599 \\   
2,1,1 &  0.441531 \\  
2,2,0 &  0.449352 \\   
3,0,0 &  0.450372 \\  
3,1,0 &  0.454416 \\   
4,0,0 &  0.464884 \\   
$\infty$ & \hspace{6.5mm}0.505462\cite{Watson,Joyce} \\ 
\end{tabular}
\caption{Numerical values of $R(l_1,l_2,l_3)$ in units of $R$ 
for a simple cubic lattice.}
\label{table-1}
\end{table} 
Our results are in agreement 
with those given by Atkinson\cite{Atkinson}.

It is worth mentioning some useful references for the lattice Green's function 
of other three dimensional lattices, such as body centered and face centered. 
The exact values of $G(1;0,0,0)$ 
as well as other exact results can be found 
in Joyce's paper\cite{Joyce2} for a body centered cubic lattice.  
Morita \cite{Morita} gave the recurrence relations of the lattice 
Green's function for a body centered cubic lattice.
Inoue\cite{Inoue} derived the recurrence relations for a face centered cubic 
lattice and obtained the exact values of the lattice
Green's function at some lattice points. 
Based on the results given in these papers one can calculate 
the resistance between arbitrary nodes for cubic lattices. 

Finally we would like to point out that 
in Eq.\ (\ref{vanHove}) the resistance is related to the energy 
dependent Green's function at energy $E=3$. However, it is known 
that the density of states at this energy  corresponds to one type 
of the van Hove singularities \cite{Ziman,Ashcroft}. Therefore, 
the resistance for a simple cubic lattice is related to 
the Green's function of 
the tight-binding Hamiltonian at the value of the energy at which 
the density of states has a van Hove type singularity. 
In the general case of a $d$ dimensional hypercube the Green's function of 
the tight-binding Hamiltonian is infinite 
at $E=d$ for $d=2$  (logarithmically divergent) 
while it has a finite value for $d=3$ as it was shown above. In the latter
case the derivative of the Green's function with respect to $E$ is singular.

\section{Rectangular lattice}  \label{sec:rect} 

In this section we shall calculate the resistance of a rectangular 
lattice in which the resistance of each edge is proportional 
to its length. 
Consider a rectangular lattice with unit vectors 
${\bf a}_1$ and ${\bf a}_2$ and introduce parameter 
$p=\left|{\bf a}_1\right|/\left|{\bf a}_2\right|.$ 
Let $R$ be the resistance of the edge along the direction of ${\bf a}_2$. 
To find the resistance between the origin and site ${\bf r}_0$
assume a current $I$ enters at the origin and exits at site ${\bf r}_0$.
We denote the potential at lattice point ${\bf r}$ by $V({\bf r})$. 
Then, according to Ohm's and Kirchhoff's laws, we may write 
\begin{equation}
\triangle_{{\rm r}}^{{\rm rect}}
V({\bf r}) = - I \left(\delta_{{\bf r},0} -
\delta_{{\bf r},{\bf r}_0} \right) R,
\label{aram-1-rect}
\end{equation}
where the `rectangular' Laplacian is defined by 
\begin{equation}
\triangle_{{\rm r}}^{{\rm rect}}
V({\bf r}) =\frac{V({\bf r}+{\bf a}_1)-V({\bf r})}{p}+
\frac{V({\bf r}-{\bf a}_1)-V({\bf r})}{p}+
V({\bf r}+{\bf a}_2)-V({\bf r})+
V({\bf r}-{\bf a}_2)-V({\bf r}).
\label{rect-Laplacian}
\end{equation}
Then the equation for the lattice Green's function  corresponding to 
the `rectangular' Laplace equation is 
\begin{equation}
\triangle_{{\rm r}^\prime}^{{\rm rect}}
G({\bf r}-{\bf r}^\prime) =  -
\delta_{{\bf r},{\bf r}^\prime}. 
\label{Green-rect}
\end{equation}
The Green's function can be calculated in a way similar to the  
hypercubic case in Sec.\ \ref{sec:hyper}. We have
\begin{equation}
G({\bf r}) = v_0\int_{{\bf k} \in {\rm BZ}} \, 
\frac{d^d {\bf k}}{{(2\pi)}^d}\, 
\frac{e^{i{\bf k}{\bf r}}}{\varepsilon({\bf k})},
\label{G-rect-veg}
\end{equation}
where $v_0=pa^2$ is the area of the unit cell, and the Brillouin zone is 
a rectangle with sides $2\pi/\left|{\bf a}_1\right|$ 
and $2\pi/\left|{\bf a}_2\right|$ along the directions of ${\bf a}_1$ and 
${\bf a}_2$, respectively, and 
\begin{equation}
\varepsilon({\bf k})= 2\left[\frac{1}{p}
\left(1-\cos {\bf k}{\bf a}_1\right) + 
1 - \cos {\bf k}{\bf a}_2  \right].
\end{equation}
Using Eq.\ (\ref{R-veg}), the resistance between the origin and 
lattice point ${\bf r}_0=l_1{\bf a}_1 +l_2 {\bf a}_2$ for a given $p$ 
is 
\begin{eqnarray}
R(p;l_1,l_2) & = & R \int_{-\pi}^{\pi} \frac{d x_1}{2\pi}
\int_{-\pi}^{\pi} \frac{d x_2}{2\pi}\, \, \,  
\frac{1-e^{il_1 x_1+il_2 x_2}}
{\frac{1}{p}\left(1-\cos x_1\right) +1- \cos x_2}
\label{R-rect-a} \\[1ex]
& = & R \int_{-\pi}^{\pi} \frac{d x_1}{2\pi}
\int_{-\pi}^{\pi} \frac{d x_2}{2\pi}\, \, \,  
\frac{1-\cos\left(l_1 x_1+l_2 x_2\right)}
{\frac{1}{p}\left(1-\cos x_1\right) +1- \cos x_2}.
\label{R-rect-b}
\end{eqnarray}
The integral over $x_2$ can be evaluated similarly to a square
lattice (see Appendix \ref{appendix-1}). We find 
\begin{equation}
R(p;l_1,l_2)= R \int_{0}^{\pi} \frac{d x}{\pi} \,\,
\frac{1-e^{-\left|l_2\right|s}\cos l_1x}{\sinh s},
\label{R-rect-one_int}
\end{equation}
where
\begin{equation}
\cosh s = 1+\frac{1}{p}-\frac{1}{p}\cos x.
\end{equation}
Note that $l_1$ and $l_2$ are interchanged here compared to  
Eq.\ (\ref{R-d2-one_int}).

\begin{figure}[hbt]
{\centerline{\leavevmode \epsfxsize=8.5cm \epsffile{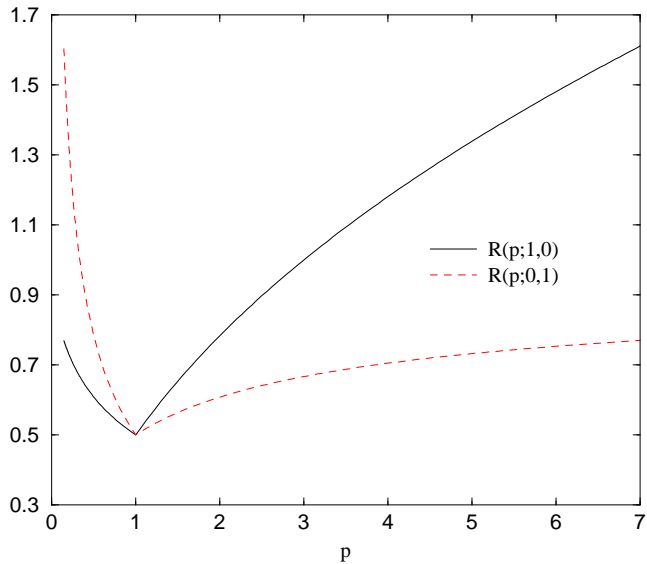 }}}
\caption{The resistances $R(p;1,0)$ and $R(p;0,1)$ in units of $R$ 
as functions of $p$ for a rectangular lattice with 
$p=\left|{\bf a}_1\right|/\left|{\bf a}_2\right|.$
\label{R-gorbe-tegla}}
\end{figure}

Now it is interesting to see how the resistance between adjacent sites
varies with $p=\left|{\bf a}_1\right|/\left|{\bf a}_2\right|.$ 
The resistance between nearest neighbors is different 
along the $x$ and the $y$ axis. 
Thus, unlike in the case of a square lattice, $R(p;1,0) \neq R(p;0,1)$ for
a rectangular lattice except for the trivial case $p=1$ (square lattice).
No sum rule exists for $R(p;1,0)+ R(p;0,1)$ as in the case of a square 
lattice. 
In Fig.\ \ref{R-gorbe-tegla} the resistances $R(p;1,0)$ and $R(p;0,1)$
are plotted as functions of $p$.
From Fig.\ \ref{R-gorbe-tegla} one can see that resistance 
$R(p;1,0)$ increases with increasing $p$. It can be shown that 
$R(p;1,0) \rightarrow \infty$ as $p\rightarrow \infty$.
This is physically clear since the lattice constant along the $x$ 
axis increases resulting in an increasing resistance of each of 
the segments in this direction. On the other hand along the $y$ axis  
a saturation of $R(p;0,1)$ can be seen, which is not obvious at all. 
Expanding the integral in the expression of $R(p;0,1)$ in powers of $p$, 
for large $p$ we get $R(p;0,1) \approx R(1-2/\pi p^{-1/2}+ O(p^{-3/2}))$. 
Thus, $R(p;0,1) \rightarrow R$ when $p\rightarrow \infty$.
Some additional results are presented for the energy dependent 
lattice Green's function for a rectangular lattice in the papers of 
Morita et.\ al., \cite{Morita-Hori} and Katsura et.\ al
\cite{Katsura-2}.

\section{Triangular lattice}  \label{sec:triang} 

In this section we calculate the resistance in a triangular 
lattice in which the resistance of each edge is identical, say $R$. 
First we consider a triangular lattice with unit vectors  
${\bf a}_1$ and ${\bf a}_2$, and with a lattice constant 
$a=\left|{\bf a}_1\right|=\left|{\bf a}_2\right|$. 
We choose  ${\bf a}_1$ and ${\bf a}_2$ such that the angle between 
them is $120^\circ$.
We introduce a third vector by 
${\bf a}_3=-({\bf a}_1+{\bf a}_2)$. The vectors drawn from each lattice
point to its 6 nearest neighbors are 
$\pm{\bf a}_1,\pm{\bf a}_2,\pm{\bf a}_3$. Assume that a current $I$ enters 
at the origin and exits at site ${\bf r}_0$, and that the potential 
at site ${\bf r}$ is $V({\bf r})$.
Again, from Ohm's and Kirchhoff's laws, we find 
\begin{equation}
\triangle_{{\rm r}}^{{\rm triang}}
V({\bf r}) = - I \left(\delta_{{\bf r},0} -
\delta_{{\bf r},{\bf r}_0} \right) R,
\label{aram-1-triang}
\end{equation}
where the `triangular' Laplacian is defined by
\begin{equation}
\triangle_{{\rm r}}^{{\rm triang}}
V({\bf r}) =\sum_{i=1}^3 \left[V({\bf r}-{\bf a}_i)-2V({\bf r})+
V({\bf r}+{\bf a}_i)\right].
\label{triang-Laplacian}
\end{equation}
It is important to note that the triangular Laplacian 
$\triangle_{{\rm r}}^{{\rm triang}}$ defined above is $2/(3a^2)$ times that 
used for solving the Laplace equation with the finite-difference method on
a triangular lattice. The factor $2/(3a^2)$ ensures that 
the lattice Laplacian in the finite-difference method yields the correct 
form of the Laplacian in the continuum limit ($a\rightarrow 0$). 
However, in our case this limit does not exist since there are also 
connections (resistors) between adjacent lattice points along the 
direction of the vector ${\bf a}_3$. 

The equation for the lattice Green's function  corresponding to 
the `triangular' Laplace equation is given by  
\begin{equation}
\triangle_{{\rm r}^\prime}^{{\rm triang}}
G({\bf r}-{\bf r}^\prime) =  -
\delta_{{\bf r},{\bf r}^\prime}. 
\label{Green-triang}
\end{equation}
The Green's function can be calculated in the same way as in the 
hypercubic case. 
We have
\begin{equation}
G({\bf r}) = v_0\int_{{\bf k} \in {\rm BZ}} \, 
\frac{d^2 {\bf k}}{{(2\pi)}^2}\, 
\frac{e^{i{\bf k}{\bf r}}}{2\sum_{i=1}^3
\left(1-\cos {\bf k}{\bf a}_i\right)},
\label{G-triang-veg}
\end{equation}
where $v_0=\sqrt{3}/2a^2$ is the area of the unit cell, and 
the vector ${\bf k}$ given in Eq.\ (\ref{k-vector}) runs over 
the Brillouin zone of the triangular lattice, which 
is a regular hexagon\cite{Lubensky}.
 
The resistance between the origin and site ${\bf r}_0$ can be 
obtained from Eq.\ (\ref{R-veg}) with the lattice Green's function 
for a triangular lattice and it yields
\begin{equation}
R({\bf r}_0) = Rv_0\int_{{\bf k} \in {\rm BZ}} \, 
\frac{d^2 {\bf k}}{{(2\pi)}^2}\, 
\frac{1-e^{i{\bf k}{\bf r}_0}}{\sum_{i=1}^3 
\left(1-\cos {\bf k}{\bf a}_i\right)} = 
Rv_0\int_{{\bf k} \in {\rm BZ}} \, 
\frac{d^2 {\bf k}}{{(2\pi)}^2}\, 
\frac{1-\cos{\bf k}{\bf r}_0}{\sum_{i=1}^3 
\left(1-\cos {\bf k}{\bf a}_i\right)}.
\label{R-triang-veg}
\end{equation}
Note that the factor $2$ dropped out from the denominator.

Using Eq.\ (\ref{R-triang-veg}) we can easily find the resistance between 
two adjacent lattice sites. 
For symmetry reasons it is clear that 
$R({\bf a}_1)=R({\bf a}_2)=R({\bf a}_3)$. On the other hand 
\begin{equation}
\sum_{i=1}^3 R({\bf a}_i) 
= Rv_0\int_{{\bf k} \in {\rm BZ}} \, \frac{d^2 {\bf k}}{{(2\pi)}^2}\, 
=R.
\end{equation}
In the last step we have made use of the fact that the volume of the 
Brillouin zone is $1/v_0$. Hence, the resistance between two adjacent 
lattice sites is $R({\bf a}_1)=R/3$. 

It is worth mentioning that without changing the resistance between two 
arbitrary lattice points one can transform the triangular lattice 
to a square lattice in which there are also resistors between the end 
points of one of the diagonals of each square.    
This topologically equivalent lattice is more suitable for evaluating 
the necessary integrals over the Brillouin zone since the Brillouin zone 
becomes a square. The same transformation was used by 
Atkinson et.\ al.\ in their paper\cite{Atkinson}.  
If we choose ${\bf a}_1=(1,0)$ and ${\bf a}_1=(0,1)$, then the resistance
between the origin and lattice point ${\bf r}_0=n{\bf a}_1+m{\bf a}_2$ is 
given by 
\begin{equation}
R(n,m)= R \int_{-\pi}^{\pi} \frac{d x}{2\pi}
\int_{-\pi}^{\pi} \frac{d y}{2\pi}\, \, \,  
\frac{1-e^{inx+imy}}{3-\cos x - \cos y - \cos(x+y)}.
\label{R-triang-veg-trafo}
\end{equation}
In Appendix \ref{appendix-3} we perform one integral in 
Eq.\ (\ref{R-triang-veg-trafo}) using the method of 
residues\cite{Arfken} in a similar way to Appendix \ref{appendix-1}. 
After some algebra it is easy to see that the result given in 
Eq.\ (\ref{triang-R-egyint-veg-app}) 
is in agreement with Atkinson's result\cite{Atkinson}: 
\begin{equation}
R(n,m)= R \int_{0}^{\pi/2} \frac{d x}{\pi} \,\,
\frac{1-e^{-\left|n-m\right|s}\cos (n+m)x}{\sinh s \, \cos x},
\label{R-triang-one_int}
\end{equation}
where $\cosh s=2\sec x  -\cos x.$ 

For $n=m$ in Eq.\ (\ref{R-triang-one_int}) we have 
\begin{equation}
R(n,n)=2R \int_{0}^{\pi/2} \frac{d x}{\pi} \,\,
\frac{1-\cos 2nx}
{\sqrt{{\left(3-\cos 2x\right)}^2 -4\cos^2 x}}.
\end{equation}
Evaluating this expression by means of the program Maple  
we obtained the same results for $n=m$ as Atkinson\cite{Atkinson}. 
Like in the case of a square lattice we believe that similar recurrence 
formulas exist for a triangular lattice but further work is necessary 
along this line. Similarly, it would be interesting to find the 
asymptotic form of the resistance as the separation between the 
two nodes tends to infinity. According to some preliminary work 
the resistance is again logarithmically divergent.

\section{Honeycomb lattice}  \label{sec:honey}

In this section we calculate the resistances in an infinite honeycomb
lattice of resistors. Atkinson and Steenwijk\cite{Atkinson} 
have studied this lattice structure exploiting the fact that 
the hexagonal lattice can be constructed from the triangular lattice 
by the application of the so-called $\Delta-Y$ transformation\cite{Duffin}.  
One advantage of our Green's function method is that we do not use
such  a $\Delta-Y$ transformation, specific only for the honeycomb lattice,
therefore our method can also be used in a straightforward manner 
for other lattice structures. As we shall see later, in the honeycomb 
lattice each unit cell contains $two$ lattice points, which is the main 
structural difference from the triangular lattice. Including more than 
one lattice point in the unit cell, the method outlined in this section 
can be viewed as a generalization of the Green's function method discussed 
in the previous sections.     

\begin{figure}[hbt]
{\centerline{\leavevmode \epsfxsize=8.5cm \epsffile{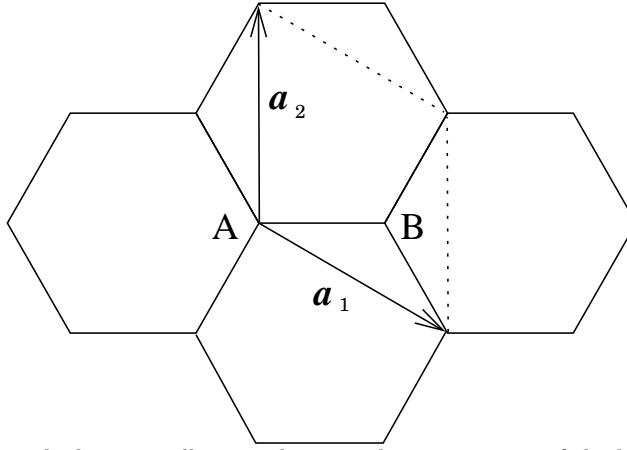 }}}
\caption{The honeycomb lattice with the unit cell. ${\bf a}_1$ and 
${\bf a}_2$ are the unit vectors of the lattice. Each unit cell contains
two types of lattice points, $A$ and $B$.
\label{hatszog-abra}}
\end{figure}

We assume that all the resistors have the same 
resistance $R$. The lattice structure and the unit cell are shown 
in Fig.\ \ref{hatszog-abra}. The angle between ${\bf a}_1$ and 
${\bf a}_2$ is $120^\circ$, and $|{\bf a}_1|=|{\bf a}_2|=\sqrt{3}a$, where
$a$ is the length of the edges of the hexagons.
There are two types of lattice points in each unit cell denoted by $A$ 
and $B$ in Fig.\ \ref{hatszog-abra}.
From now on the position of the lattice points $A$ and $B$ will be given by
the position of the unit cell in which they are located. 
Assume the origin is at one of the lattice points $A$, and then the position 
of a unit cell can be specified by the position vector 
${\bf r}=n{\bf a}_1+m{\bf a}_2$, 
where $n,m$ are arbitrary integers. We denote the potential and the current 
in one of the unit cells by $V_A({\bf r})$ and $V_B({\bf r})$, 
and $I_A({\bf r})$ and $I_B({\bf r})$, respectively, where subscripts 
$A$ and $B$ refer to the corresponding lattice points. Owing to  
Ohm's and Kirchhoff's laws the currents $I_A({\bf r})$ and 
$I_B({\bf r})$ in the unit cell specified by ${\bf r}$ satisfy 
the following equations: 
\begin{eqnarray}
I_A({\bf r}) & = & \frac{V_A({\bf r})-V_B({\bf r})}{R}
+\frac{V_A({\bf r})-V_B({\bf r}-{\bf a}_1)}{R}+
\frac{V_A({\bf r})-V_B({\bf r}-{\bf a}_1-{\bf a}_2)}{R},
\nonumber \\[1ex]
I_B({\bf r}) & = & \frac{V_B({\bf r})-V_A({\bf r})}{R}
+\frac{V_B({\bf r})-V_A({\bf r}+{\bf a}_1)}{R}+
\frac{V_B({\bf r})-V_A({\bf r}+{\bf a}_1+{\bf a}_2)}{R}.
\label{aram-honey}
\end{eqnarray}

Assuming periodic boundary conditions again the potential 
$V_A({\bf r})$ can be given by its Fourier transform
\begin{equation} 
V_A({\bf r}) =\frac{1}{N}\sum_{{\bf k} \in {\rm BZ}} 
V_A({\bf k})\,  e^{i{\bf k}{\bf r}},
\label{hatsz-fourier-pot}
\end{equation} 
where $N$ is the number of unit cells, and analogous 
expressions are valid for $V_B({\bf r}), I_A({\bf r})$ and 
$I_B({\bf r})$. 
Here ${\bf k}$ is in the Brillouin zone\cite{Ashcroft,Ziman,Kittel}. 
Thus, we may rewrite Eq.\ (\ref{aram-honey}) as  
\begin{equation}
{\bf L}({\bf k})\left [ \begin{array} {c} 
V_A({\bf k})  \\
V_B({\bf k})   \end{array} \right ]
=-R\left [ \begin{array} {c} 
I_A({\bf k})  \\
I_A({\bf k})   \end{array} \right ],
\label{fourier}
\end{equation}
where
\begin{equation}
{\bf L}({\bf k}) = \left ( \begin{array} {cc} 
-3  & h^* \\
h & -3  \end{array}  \right ), 
\end{equation}
and 
\begin{equation}
h=1+e^{i{\bf k}{\bf a}_1}+e^{i{\bf k}\left({\bf a}_1+{\bf a}_2\right)}.
\label{kis_h-def}
\end{equation}
Equation (\ref{fourier}) is indeed the Fourier transform of the
Poisson-like equation that determines the potentials for a given current
distribution. However, in this case the Laplace operator is a 2x2 matrix. 
In ${\bf r}$-represantion the analogous equation for a hypercubic lattice 
was given in Eq.\ (\ref{aram-2}). 

The equation for the Fourier transform of the Green's function (which is 
a 2x2 matrix for a honeycomb lattice, too) can be defined analogously to 
a hypercubic lattice in Eq.\ (\ref{Green-1}):
\begin{equation}
{\bf L}({\bf k}){\bf G}({\bf k}) = -1.
\label{Green-hatsz-def}
\end{equation}

The solution of Eq.\ (\ref{Green-hatsz-def})  for the Green's function is 
\begin{equation}
{\bf G}({\bf k}) = \frac{1}{9-{\left|h\right|}^2}
\left ( \begin{array} {cc} 
3  & h^* \\
h & 3  \end{array}  \right ), 
\label{Green-hatsz}
\end{equation}
where $9-{\left|h\right|}^2=
2\left(3-\cos {\bf k}{\bf a}_1-\cos {\bf k}{\bf a}_2-
\cos {\bf k}{\bf a}_3 \right)$ and we have  
introduced a third vector, 
${\bf a}_3=-\left({\bf a}_1+{\bf a}_2\right)$.

There are four types of resistance.   
We denote the resistance between a lattice point $A$ as the origin and 
site ${\bf r}_0=n{\bf a}_1+m{\bf a}_2$ (which is an $A$-type site) 
by $R_{AA}({\bf r}_0)$, 
while the resistance between the origin and site  
${\bf r}_0+(2{\bf a}_1+{\bf a}_2)/3$ (which is a $B$-type site in the 
unit cell at ${\bf r}_0$) is denoted by 
$R_{AB}({\bf r}_0)$. For symmetry reasons, it follows that for the other 
two types of resistance: $R_{BB}({\bf r}_0)=R_{AA}({\bf r}_0)$ and 
$R_{BA}({\bf r}_0)=R_{AB}({\bf r}_0).$
To measure $R_{AA}({\bf r}_0)$ the current at sites $A$ and $B$ in the unit 
cell at ${\bf r}$ are 
\begin{equation}
I_A({\bf r})=I\left(\delta_{{\bf r},0} -
\delta_{{\bf r},{\bf r}_0} \right)\,\,\,\,\,
\mbox{and} \,\,\,\,\,I_B({\bf r})=0,
\label{aram-a}
\end{equation}
while for measuring $R_{AB}({\bf r}_0)$ we have
\begin{equation}
I_A({\bf r})=I \, \delta_{{\bf r},0}\,\,\,\,\,
\mbox{and}\,\,\,\,\, I_B({\bf r})= -I \, \delta_{{\bf r},{\bf r}_0}.
\label{aram-b}
\end{equation}
Thus
\begin{eqnarray}
R_{AA}({\bf r}_0) & = & \frac{V_A({\bf r}={\bf 0})-V_A({\bf r}_0)}{I}.
\label{R-a}
\\[1ex] 
R_{AB}({\bf r}_0) & = & \frac{V_A({\bf r}={\bf 0})-V_B({\bf r}_0)}{I}.
\label{R-b}
\end{eqnarray}

First, consider the resistance $R_{AA}({\bf r}_0)$.  
From Eqs.\ (\ref{fourier}), 
(\ref{Green-hatsz-def})--(\ref{aram-a}) we obtain
\begin{equation}
V_A({\bf k})=IR\,G_{11}({\bf k})\left(1-e^{-i{\bf k}{\bf r}_0} \right).
\end{equation}
Hence, using  Eq.\ (\ref{hatsz-fourier-pot}) we can obtain the 
${\bf r}$-dependence of  the potential $V_A({\bf r})$ and then 
Eq.\ (\ref{R-a})  yields
\begin{equation}
R_{AA}({\bf r}_0)=R\frac{2}{N}\sum_{{\bf k}\in {\rm BZ}} G_{11}({\bf k})
\left(1-\cos {\bf k}{\bf r}_0\right)=
R\frac{3}{N}\sum_{{\bf k}\in {\rm BZ}}
\frac{1-\cos {\bf k}{\bf r}_0}
{3-\cos {\bf k}{\bf a}_1-\cos {\bf k}{\bf a}_2-
\cos {\bf k}{\bf a}_3}.
\label{R-a-veg-sum}
\end{equation}
Using Eq.\ (\ref{sum-int}), the summation over ${\bf k}$ can be 
substituted by an integral and we obtain 
\begin{equation}
R_{AA}({\bf r}_0) = 3 Rv_0\int_{{\bf k} \in {\rm BZ}} \, 
\frac{d^2 {\bf k}}{{(2\pi)}^2}\,
\frac{1-\cos {\bf k}{\bf r}_0}
{3-\cos {\bf k}{\bf a}_1-\cos {\bf k}{\bf a}_2-
\cos {\bf k}{\bf a}_3},
\label{R-a-veg}
\end{equation}
where $v_0=3\sqrt{3}a^2$ is the area of the unit cell. One can see
that the same expression was found for a triangular lattice except for the
factor $3v_0$ here [see Eq.\ (\ref{R-triang-veg})]. Note that although 
the area of the unit cell $v_0$ is 6 times bigger than that for the 
triangular lattice, the area of the Brillouin zone is 6 times less and thus, 
the ratio of the resistances for honeycomb and triangular lattices is 3.  
Therefore, the resistance between two $A$-type nodes is 
three times the corresponding resistance in the triangular lattice. 
This again agrees with Atkinson's result. 
 
Similarly, using Eqs.\ (\ref{fourier}), (\ref{Green-hatsz-def}), 
(\ref{Green-hatsz}) and (\ref{aram-b}) we can obtain $V_A({\bf k})$ 
and $V_B({\bf k})$. 
From Eq.\ (\ref{hatsz-fourier-pot}) the potential $V_A({\bf r})$ 
can be determined, and analogously $V_B({\bf r})$. 
Finally, Eq.\ (\ref{R-b}) leads to
\begin{eqnarray}
R_{AB}({\bf r}_0) & = & R\frac{1}{N}\sum_{{\bf k}\in {\rm BZ}} 
\left(G_{11}({\bf k})+G_{22}({\bf k})
-G_{12}({\bf k})\, e^{-i{\bf k}{\bf r}_0}-
G_{21}({\bf k})\, e^{i{\bf k}{\bf r}_0}\right)
\nonumber \\[1ex]
& = &  Rv_0\int_{{\bf k} \in {\rm BZ}} \, 
\frac{d^2 {\bf k}}{{(2\pi)}^2}\,
\frac{3-\cos {\bf k}{\bf r}_0-
\cos {\bf k}\left({\bf r}_0+{\bf a}_1\right)
-\cos {\bf k}\left({\bf r}_0+{\bf a}_1+{\bf a}_2\right)}
{3-\cos {\bf k}{\bf a}_1-\cos {\bf k}{\bf a}_2-
\cos {\bf k}{\bf a}_3}
\nonumber \\[1ex]
& = & \frac{1}{3}
\left[R_{AA}({\bf r}_0)
+R_{AA}({\bf r}_0+{\bf a}_1)
+R_{AA}({\bf r}_0+{\bf a}_1+{\bf a}_2) \right].
\label{R-b-veg}
\end{eqnarray}
The same result was found by Atkinson and Steenwijk.

The resistance between second nearest neighbor lattice sites may be found
from Eq.\ (\ref{R-a-veg}). For symmetry reasons  
$R_{AA}({\bf a}_1)=R_{AA}({\bf a}_2)=R_{AA}({\bf a}_3)$ and 
from  Eq.\ (\ref{R-a-veg}) we have $\sum_{i=1}^3 R_{AA}({\bf a}_i)= 3R.$ 
Thus, the resistance between second nearest neighbor lattice sites is $R.$
Using Eq.\ (\ref{R-b-veg}), the resistance for adjacent lattice sites is   
$R_{AB}({\bf 0})=1/3\left[R_{AA}({\bf 0})+R_{AA}({\bf
a}_1)+R_{AA}({\bf a}_1+{\bf a}_2)\right]=2R/3$, since obviously, $R_{AA}({\bf
0})=0$. 
We would just mention that the limiting value of the resistance 
is again infinite as the separation between nodes tends to infinity
(see p.\ 139 of Doyle's and Snell's book\cite{Doyle}).
  
\begin{figure}[hbt]
{\centerline{\leavevmode \epsfxsize=8.5cm \epsffile{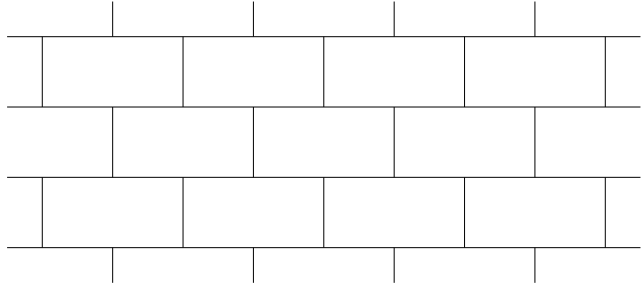 }}}
\caption{The honeycomb lattice can be transformed into a topologically 
equivalent brick-type lattice. 
\label{tegla-abra}}
\end{figure}

We note that one can transform the honeycomb lattice 
to a topologically equivalent brick-type lattice shown 
in Fig.~ \ref{tegla-abra} without changing the resistance between two 
arbitrary lattice points. 
This fact has been utilized in the case of
the triangular lattice in the previous section.  

As it can be seen from the above results, the resistance between 
arbitrary lattice points in a honeycomb lattice can
be related to the corresponding triangular lattice. This kind of relation
is the consequence of the so-called {\it duality transformation\/} 
in which the variables are transformed to Fourier transform variables 
(for more details see Chaikin's book\cite{Lubensky} pp.\ 578--584).
The dual lattice of a triangular lattice is a honeycomb lattice.
However, if the unit cell contains more than two non-equivalent lattice
points then such a connection might not be used. On the other hand, 
the Green's function method outlined in the example of the honeycomb 
lattice can still be applied straightforwardly.

\acknowledgements

The author wishes to thank  G.\ Tichy, T.\ Geszti, 
P.\ Gn\"adig, A.\ Pir\'oth, S.\ Redner and L. Glasser 
for helpful discussions.
This work was supported by the Hungarian Science Foundation OTKA  
T025866, the Hungarian Ministry of Education (FKFP 0159/1997) and OMFB
within the program ``Dynamics of Nanostructures''.

\appendix 
\section{Integration in the expression of the resistance 
for a square lattice} \label{appendix-1}

Starting from  Eq.\ (\ref{R-d2-a}) for the resistance $R(n,m)$ we have
\begin{equation}
R(n,m)= R\int_{-\pi}^{\pi} \, \frac{d y}{2\pi} \,\, I(y),
\end{equation}
where
\begin{equation}
I(y)= \int_{-\pi}^{\pi} \, \frac{d x}{2\pi} \,\,
\frac{1-e^{i n x}e^{i m y}}{2-\cos x - \cos y}.
\label{I-y}
\end{equation}
Since $R(n,m)=R(-n,m)$, we take $n > 0$. 
To proceed further, we perform the integral in $I(y)$  using the 
method of residues. 
Introducing the complex variable 
$z=e^{ix}$ the integral $I(y)$ reads 
\begin{equation}
I(y)= -2i \oint \,\, \frac{dz}{2\pi}\,\,\,
\frac{1-z^n e^{i m y}}{2z\left(2-\cos y\right)-z^2-1},
\end{equation}
where the path of integration is the unit circle. 
The denominator has roots at $z_{+}=e^{i\alpha_{+}}$ and 
$z_{-}=e^{i\alpha_{-}}$, where $\alpha_{+}$ and $\alpha_{-}$ satisfy the 
equation  $\cos \alpha = 2-\cos y$ and $\alpha_{-}=-\alpha_{+}.$
It is clear that for $-\pi < y < \pi$ we have  $2-\cos y > 1$,  
therefore the solution for $\alpha$ is purely imaginary. 
Thus we introduce $s$ with  $\alpha_{+}= -\alpha_{-}=is$, 
where $s$ satisfies the equation
\begin{equation}
\cosh s= 2-\cos y.
\label{gyok}
\end{equation}
For $-\pi < y < \pi$ it is true that $s > 0$, so  
the two poles of the integrand in $I(y)$ are real numbers and satisfy 
the following inequalities: $z_{+}=e^{-s} < 1$ and $z_{-}=e^{s} > 1$. 
Thus the pole  $z_{+}$ is within the unit circle, while $z_{-}$ is outside.
According to the residue theorem\cite{Arfken} $I(y)= -2i 2\pi i 
\sum \mbox {residues within the unit circle}$.
Using Eq.\ (\ref{gyok}) the residue of the integrand of 
$I(y)$ at $z_{+}$ is 
\begin{equation}
\frac{1}{2\pi}
\frac{1-e^{-ns}e^{imy}}{2\left(2-\cos y\right)-2z_{+}}=
\frac{1}{2\pi}
\frac{1-e^{-ns}e^{imy}}{2\sinh s}.
\end{equation}
We obtain
\begin{equation}
I(y)=-2i2\pi i
\frac{1}{2\pi}
\frac{1-e^{-ns}e^{imy}}{2\sinh s}=
\frac{1-e^{-ns}e^{imy}}{\sinh s},
\label{I-y-veg}
\end{equation}
where $s$ satisfies Eq.\ (\ref{gyok}).
Finally, the resistance $R(n,m)$ for arbitrary integers $n,m$ becomes
\begin{equation}
R(n,m)= R\int_{-\pi}^{\pi} \, \frac{d y}{2\pi} \,\, 
\frac{1-e^{-\left|n\right|s}e^{imy}}{\sinh s}= 
R\int_{0}^{\pi} \, \frac{d y}{\pi} \,\, 
\frac{1-e^{-\left|n\right|s}\cos my}{\sinh s}.
\end{equation}
The same result was found by Venezian\cite{Venezian}.

\section{The Asymptotic form of the lattice Green's function  
for a square lattice} \label{appendix-2}

In this Appendix we derive the asymptotic form of the lattice Green's 
function for a square lattice. The lattice Green's function at site 
${\bf r}=0$ is divergent since $\varepsilon({\bf k})=0$ for ${\bf k}=0$. 
Therefore we calculate the asymptotic form of $G({\bf 0})-G({\bf r})$.
Starting from  Eq.\ (\ref{G-hypercube}) the lattice Green's function 
for site ${\bf r}=n{\bf a}_1 +m{\bf a}_2$ in a square lattice becomes
\begin{equation}
G({\bf 0})-G(n,m) = \frac{1}{2} 
\int_{-\pi}^{\pi} \, \frac{d y}{2\pi} \,\, 
\int_{-\pi}^{\pi} \, \frac{d x}{2\pi} \,\,
\frac{1-e^{i n x}e^{i m y}}{2-\cos x - \cos y}.
\end{equation}
The integral over $x$ is the same as $I(y)$ in Eq.\ (\ref{I-y}), 
so we can use the result obtained in Eq.\ (\ref{I-y-veg}):
\begin{equation}
G({\bf 0})-G(n,m) = \frac{1}{2} 
\int_{-\pi}^{\pi} \frac{d y}{2\pi} \,\,
\frac{1-e^{-\left| n\right| s}e^{imy}}{\sinh s}=
\int_{0}^{\pi} \frac{d y}{2\pi} \,\,
\frac{1-e^{-\left| n \right| s}\cos my}{\sinh s}.
\label{G-asym-1}
\end{equation}
We follow the same method for deriving the asymptotic form of the
lattice Green's function for large $m$ and $n$ as  
Venezian\cite{Venezian}.
A similar method was used in Chaikin's book\cite{Lubensky} 
(see pp.\ 295--296) in the case of 
a continuous medium in two dimensions.
We break the integral in  Eq.\ (\ref{G-asym-1}) into three parts:
\begin{equation}
G({\bf 0})-G(n,m) = I_1 + I_2 + I_3,
\label{G-asym-2}
\end{equation}
where 
\begin{eqnarray}
I_1 & = & \int_{0}^{\pi} \frac{d y}{2\pi} \,\,
\frac{1-e^{-\left|n\right|y}\cos my}{y}, 
\nonumber  \\[1ex]
I_2 & = & \int_{0}^{\pi} \frac{d y}{2\pi} \,\, 
\left(\frac{1}{\sinh s} -\frac{1}{y}\right),   
\nonumber \\[1ex] 
I_3 & = & \int_{0}^{\pi} \frac{d y}{2\pi} \,\,
\left(\frac{e^{-\left|n\right|y}\cos my}{y} - 
\frac{e^{-\left|n\right|s}\cos my}{\sinh s}\right),
\label{I-123}
\end{eqnarray}
and $s$ satisfies Eq.\ (\ref{gyok}). 
The first integral $I_1$ can be expressed by the integral exponential 
$\mbox {Ein}(z)$ \cite{Abramowitz}:
\begin{equation}
I_1 = \frac{1}{2\pi} \mbox{Re} 
\left \{ 
\int_0^{\pi}\, dy \, \frac{1-e^{\left(\left|n \right|-im\right) y}}{y}
\right \} = \frac{1}{2\pi} \mbox{Re} 
\left \{ \mbox {Ein}\left[\pi\left(\left|n \right|-im \right)\right]
\right \},
\end{equation}
where $\mbox {Ein}(z)$ is defined by 
\begin{equation}
\mbox {Ein}(z)= \int_0^z \, dt \, \frac{1-e^{-t}}{t}.
\end{equation}
For large values of its argument, 
$\mbox {Ein}(z)\approx \ln z + \gamma$, where 
$\gamma = 0.5772\dots$ is the Euler-Mascheroni
constant. Thus, for large $n$ and $m$ $I_1$ can be approximated by
\begin{equation}
\mbox I_1 \approx \frac{1}{2\pi} \left(
\ln \left|\pi\left(\left|n \right|-im \right)\right| + \gamma \right)=
\frac{1}{2\pi} \left( \ln \sqrt{n^2 +m^2} + \gamma + \ln \pi \right).
\end{equation}

Using Eq.\ (\ref{gyok}) the integral $I_2$ can be evaluated exactly:
\begin{equation}
I_2 = \int_0^{\pi} \, \frac{d y}{2\pi}
\left(\frac{1}{\sqrt{{\left(2-\cos y \right)}^2-1}} -\frac{1}{y}\right) = 
\frac{1}{2\pi}\left( \frac{\ln 8}{2} - \ln \pi \right) .
\end{equation}
 
In the integral $I_3$ the integrand is close to zero for small
values of $y$ and $s$ since 
$s \approx \sinh s \approx y$, while for larger values of $y$ and $s$ 
the exponentials are negligible, therefore $I_3 \approx 0$.

Finally, we find that the lattice Green's function for large arguments, 
$i.e.,$ $\left|{\bf r} \right|= a\sqrt{n^2 +m^2} \rightarrow \infty$ 
becomes 
\begin{equation}
G({\bf r}) = G({\bf 0}) -\frac{1}{2\pi} 
\left (\ln \frac{\left| {\bf r}\right| }{a} +\gamma + \frac{\ln 8}{2} 
\right). 
\label{G-asym-veg}
\end{equation}
The same result is quoted on page 296 of Chaikin's book\cite{Lubensky}. 
In the theory of the Kosterlitz--Thouless--Berezinkii phase transition 
\cite{KT,Kosterlitz,Berez} Kosterlitz  used the same asymptotic 
form of the Green's function for a square lattice \cite{Kosterlitz}. 

\section{Integration in the expression of the resistance 
for a triangular lattice} \label{appendix-3}

Introducing the coordinate transformations 
$x^\prime =  (x+y)/2$ and $y^\prime = (x-y)/2$, 
and using the complex variable $z=e^{iy^\prime/2}$ we have
$R(n,m)=1/2 \, R \, \int_{-\pi}^\pi dx^\prime/2\pi \,\, I(x^\prime)$ 
from Eq.\ (\ref{R-triang-veg-trafo}), where 
\begin{equation}
I(x^\prime)=2i \oint \,\, \frac{dz}{2\pi}\,\,\,
\frac{1-z^{(n-m)}\,\, e^{i (n+m) \frac{x^\prime}{2}}}
{z^2\cos \frac{x^\prime}{2} 
-z\left(3-\cos x^\prime \right) +\cos \frac{x^\prime}{2}},
\end{equation}
and the path of integration is the unit circle. The factor 1/2 
in front of $R(n,m)$ is the Jacobian  
corresponding to the transformations of variables.
Since $R(n{\bf a}_1-m{\bf a}_2)=R(m{\bf a}_1-n{\bf a}_2)$,  
we take $n-m > 0.$ 
The denominator has roots in $z_{+}=e^{i\alpha_{+}/2}$ and  
$z_{-}=e^{i\alpha_{-}/2}$ and it is easy to see that they satisfy 
equations $\alpha_{+}=-\alpha_{-}$ 
and  $2\cos (\alpha_{+}/2) =(3-\cos x^\prime)/\cos (x^\prime/2)$.
For $-\pi < x^\prime < \pi$ it is clear that $\alpha_{+}$ is purely 
imaginary. If we choose $\alpha_{+}$ such that 
${\rm Re}\, \alpha_{+} >0 $, then  $z_{+} < 1$ and  $z_{-} > 1.$ 
Thus, with $\alpha_{+}=is$, the residue of the integrand 
of $I(x^\prime)$ at $z_{+}$ 
(which is inside the unit circle, $i.e.,$ $s>0$) is 
\begin{equation}
\frac{1}{2\pi}
\frac{1-e^{-(n-m)\frac{s}{2}}\,\, e^{i (n+m) \frac{x^\prime}{2}}}
{2z_{+}\cos \frac{x^\prime}{2}
-\left(3-\cos x^\prime \right)}=
-\frac{1}{2\pi}
\frac{1-e^{-(n-m)\frac{s}{2}} \,\, e^{i (n+m) \frac{x^\prime}{2}}}
{2\sinh \frac{s}{2} \,\, \cos \frac{x^\prime}{2}},
\end{equation}
where $s$ satisfies the equation
\begin{equation}
2\cosh \frac{s}{2} =\frac{3-\cos x^\prime}{\cos \frac{x^\prime}{2}}.
\label{triang-s-eq}
\end{equation}
Finally, the resistance for arbitrary integers $n,m$ is 
\begin{equation}
R(n,m)=\frac{R}{2}\int_{-\pi}^\pi \frac{dx^\prime}{2\pi} \,\, 
\frac{1-e^{-\left|n-m\right|\frac{s}{2}} \,\, e^{i (n+m) \frac{x^\prime}{2}}}
{\sinh \frac{s}{2} \,\, \cos \frac{x^\prime}{2}}.
\label{triang-R-egyint-veg-app}
\end{equation}

\newpage


\end{document}